\documentclass[english]{article}
\usepackage{amsfonts}
\usepackage{amssymb}
\usepackage{fontenc}
\usepackage[latin1]{inputenc}
\usepackage{amsmath}
\usepackage{fontenc}
\usepackage{babel}

 
\input{tcilatex}

\begin{document}

\title{\textbf{On "Schwinger Mechanism for Gluon Pair Production in the
Presence of Arbitrary Time Dependent Chromo-Electric Field" }}
\author{S.P. Gavrilov\thanks{%
Department of General and Experimental Physics, Herzen State Pedagogical
University of Russia, Moyka emb. 48, 191186 St. Petersburg, Russia; e-mail:
gavrilovsergeyp@yahoo.com} and D.M. Gitman \thanks{%
Institute of Physics, University of São Paulo, CP 66318, CEP 05315-970 São
Paulo, SP, Brazil; e-mail: gitman@dfn.if.usp.br}}
\maketitle

\begin{abstract}
Recently the paper "Schwinger Mechanism for Gluon Pair Production in the
Presence of Arbitrary Time Dependent Chromo-Electric Field" by G. C. Nayak
was published [Eur. Phys. J. C \textbf{59}, 715 (2009); arXiv:0708.2430].
Its aim is to obtain an exact expression for the probability of
non-perturbative gluon pair production per unit time per unit volume and per
unit transverse momentum in an arbitrary time-dependent chromo-electric
background field. We believe that the obtained expression is open to
question. We demonstrate its inconsistency on some well-known examples. We
think that this is a consequence of using the so-called "shift theorem"
[arXiv:hep-th/0609192] in deriving the expression for the probability. We
make some critical comments on the theorem and its applicability to the
problem in question.

\end{abstract}

I. In a recent paper \cite{Nepjc09}, an expression for $\frac{dW}{%
d^{4}xd^{2}p_{T}}$, the probability of non-perturbative gluon pair
production per unit time per unit volume and per unit transverse momentum in
an arbitrary time-dependent chromo-electric background field $E^{a}\left(
t\right) $ with arbitrary color index $a=\overline{1,8}$ in $SU(3)$, was
represented. We believe that the obtained expression is open to question.

The main assertion of the paper \cite{Nepjc09} reads: "We find that the
exact result for non-perturbative gluon pair production is independent of
all the time derivatives $d^{n}E^{a}/dt^{n}$, where $n\in \mathbb{N}$, and
has the same functional dependence on two Casimir invariants $%
C_{1}=E^{a}\left( t\right) E^{a}\left( t\right) $ and $C_{2}=\left[
d_{abc}E^{a}\left( t\right) E^{b}\left( t\right) E^{c}\left( t\right) \right]
^{2}$ as the constant chromo-electric field $E^{a}$ result with the
replacement: $E^{a}\rightarrow E^{a}\left( t\right) $." The result is
presented by Eqs. (1-2) from \cite{Nepjc09} as follows:%
\begin{equation}
\frac{dW_{g\left( \bar{g}\right) }}{dtd^{3}xd^{2}p_{T}}=\frac{1}{4\pi ^{3}}%
\sum_{j=1}^{3}\left\vert g\Lambda _{j}\left( t\right) \right\vert \ln \left[
1+\exp \left( -\frac{\pi p_{T}^{2}}{\left\vert g\Lambda _{j}\left( t\right)
\right\vert }\right) \right] ,  \label{1}
\end{equation}%
where $\Lambda _{j}\left( t\right) $ is expressed via Casimir invariants $%
C_{1}$ and $C_{2}$. Here $W_{g\left( \bar{g}\right) }=2\func{Im}S^{\left(
1\right) }$ is imaginary part of the one loop (or gluon pair) effective
action $S^{\left( 1\right) }.$ The field $E^{a}\left( t\right) $ acts along
the $z$-axis, the potential $A_{3}^{a}=0,$ such that $A_{\mu }^{a}=-\delta
_{\mu 0}E^{a}\left( t\right) z.$

First, one can see that eq. (\ref{1}) is in contradiction with well-known
and well-justified results obtained previously. To this end it is enough to
consider an Abelian-like background, in which the chromo-electric field has
only one nonzero component, $E^{a}\left( t\right) =\delta _{a0}E\left(
t\right) $. Then $C_{1}=\left[ E\left( t\right) \right] ^{2}$ and $C_{2}=0$,
which implies $\Lambda _{1}\left( t\right) =\left\vert E\left( t\right)
\right\vert $ and $\Lambda _{2,3}\left( t\right) =\left\vert E\left(
t\right) \right\vert /2$. If so, then it follows from eq. (\ref{1}) that $%
W_{g\left( \bar{g}\right) }$ depends in the general case (in an arbitrary
field $E\left( t\right) $) on the time $t$. In turn, this implies (and this
is an exact result according to the author) that the probability \ (\ref{1})
is a local quantity of which the value at the time-moment $t$ is determined
by the electric $E\left( t\right) $ at the same time moment. In the case
under consideration (Abelian-like background) calculations in QCD are quite
analogous to that in QED. That is why one can consider similar calculations
in QED, where they can be performed exactly, e.g., for the electric field of
the form $E(t)=E_{0}\cosh ^{-2}\left( t/\alpha \right) ,$ where $E_{0}$ and $%
\alpha $ are some positive constants. Such a field switches on and off
adiabatically at $x^{0}\rightarrow \pm \infty $. First, differential mean
numbers $\aleph _{m}$ of particles created by such a field were found in 
\cite{NarN70}; see also \cite{FraGi81,GMM94}). The explicit form of $\aleph
_{m}$ is not important for the further consideration. What we only need to
stress that these quantities depend essentially on $\alpha ,$ they have an
essentially different form in a constant electric field, and they do not
depend on time $t$ at all. The latter fact is a general property of $\aleph
_{m}$ for any external field, since they are defined via an inner product of
solutions of a corresponding wave equation conserved in time. Using the
general formulation (see \cite{FraGi81,GMM94}), one can see that, in an
arbitrary time-dependent but uniform external field, the probability $P^{v}$
of there occuring no actual pair creation in the history of the field in the
volume $V$ can be represented as 
\begin{equation}
P^{v}=\left\vert c_{v}\right\vert ^{2}=\exp \{-2\func{Im}S^{\left( 1\right)
}\}=\exp \left\{ \kappa \sum_{m}\ln \left( 1-\kappa \aleph _{m}\right)
\right\} \,,  \label{vst9}
\end{equation}%
where $\kappa =-1$ for bosons and $\kappa =+1$ for fermions, and the
summation over $m$ implies an integration over momenta and a summation over
some discrete quantum numbers. Here $c_{v}=\langle 0,out|0,in\rangle =\exp
\left( iS^{\left( 1\right) }\right) $ is the transition amplitude from the
initial vacuum $|0,in\rangle $ to the final vacuum $|0,out\rangle $. Passing
from the summation over momenta to the integration, $\sum_{\mathbf{p}}\ldots
\rightarrow \left( 2\pi \right) ^{-3}V\int d\mathbf{p\ldots }$, there
appears a natural volume dependence for the uniform external field in (\ref%
{vst9}), $\func{Im}S^{\left( 1\right) }\sim V$. However, a time dependence
cannot appear in such a way in (\ref{vst9}) (explicit expressions can be
found in \cite{GavG96a,DunH98}). That is why the quantity $\frac{dW_{g\left( 
\bar{g}\right) }}{dtd^{3}xd^{2}p_{T}}$ makes no sense. The quantities $%
\aleph _{m}$ can also be found exactly in the case when the electric field
has an exponential fall-off, $E(x^{0})=E\exp \left( -x^{0}/\alpha \right) $ 
\cite{Spo82}, and in the case of a periodic alternating electric field \cite%
{NarNik74}. Then one can come to similar conclusions.

Likely, an illusion that the quantity $\func{Im}S^{\left( 1\right) }$
depends on time is related to the well-known Schwinger formula $\func{Im}%
S^{\left( 1\right) }\sim VT$ (see \cite{S51}), where, however, $T$ is a
total effective time of a (quasi)constant electric field action. Such a
result holds only for a quasiconstant electric field that switches on at $%
-T/2$ and switches off at $T/2,$ being quasiconstant for the total time $T$
of the field action. If the time $T$ is sufficiently large ($\sqrt{%
\left\vert qE\right\vert }T\gg 1$ for massless particles) then $\aleph _{m}$
is constant for sufficiently large region of $p_{3}$ and the integral over $%
p_{3}$ turns out to be proportional to $T$, $\int dp_{3}\aleph _{m}$ $\sim
\left\vert qE\right\vert T\aleph _{m}$; see \cite{GavG96a}. We stress that
in this case the quantity $\func{Im}S^{\left( 1\right) }$ depends on the
total time $T$ of the field action, this quantity is a global one, and $d%
\func{Im}S^{\left( 1\right) }/dT$ is constant. One ought also to mention
that interpretation of the quantity $\frac{dW}{d^{4}xd^{2}p_{T}}$ (given in 
\cite{Nepjc09}) as the probability is inaccurate in the general case (the
quantity $2\func{Im}S^{\left( 1\right) }$ can be interpreted as the total
probability of pair creation only if the WKB approximation is applicable;
see, e.g., Sect. 3.2 \cite{GavGT06}).

Thus, one can see that something is wrong with the alleged exact result (\ref%
{1}). Checking the derivation of eq. (\ref{1}), given in \cite{Nepjc09}, we
have discovered that the transformation of the effective action $S^{\left(
1\right) }$ from Eq.(36) to Eq.(37) is erroneous. This makes the result (\ref%
{1}) erroneous. Doing this transformation, authors of \cite{Nepjc09} refer
to the article \cite{CNshift06}, where such a transformation (named a "shift
theorem") was justified (see a comment on the shift theorem below).

II. The transformation from Eq.(36) to Eq.(37) given in \cite{Nepjc09},
repeats, in fact, the main assertion of the shift theorem, which in its
original form in \cite{CNshift06} is not related to peculiarities of the
SU(3) group. One can see that this transformation acts on the matrix element 
\begin{equation}
\int_{-\infty }^{+\infty }dz\left\langle z\left\vert e^{-is\left[ \left(
i\partial _{t}-g\Lambda \left( t\right) z\right) ^{2}-\hat{p}%
_{z}^{2}-2ig\lambda _{l}\Lambda \left( t\right) \right] }\right\vert
z\right\rangle  \label{3}
\end{equation}%
only, at fixed values of other variables, including color indices. Here $%
\lambda _{l}$ are eigenvalues of the Lorentz matrix, $\left( \lambda
_{1},\lambda _{2},\lambda _{3},\lambda _{4}\right) =\left( 1,0,-1,0\right) $%
, $\Lambda ^{ab}\left( t\right) =if^{abc}E^{c}\left( t\right) $. Denoting, $%
g\Lambda \left( t\right) =a\left( t\right) $, $-2ig\lambda _{l}\Lambda
\left( t\right) =c\left( t\right) $, and $-\hat{p}_{z}^{2}=b\left( \partial
_{z}\right) $, one can see that the transformation from Eq.(36) to Eq.(37)
of \cite{Nepjc09} is equivalent to the assertion of the shift theorem.

Let us denote coordinate operators by $\hat{z}$ and $\hat{t}$, and their
eigenvalues by $z$ and $t$, respectively, $\hat{z}\left\vert z\right\rangle
=z\left\vert z\right\rangle $, $\hat{t}\left\vert t\right\rangle
=t\left\vert t\right\rangle $, such that $\left\langle z\right. \left\vert
z^{\prime }\right\rangle =\delta \left( z-z^{\prime }\right) $, $%
\left\langle t\right. \left\vert t^{\prime }\right\rangle =\delta \left(
t-t^{\prime }\right) $. Then the main assertion (4) of the shift theorem
from \cite{CNshift06} is equivalent to the validity of the following
relation: 
\begin{eqnarray}
&&\int_{-\infty }^{+\infty }dzf_{1}\left( \hat{t}\right) \left\langle
z\left\vert e^{-u\left[ \left( \hat{P}_{t}-a\left( \hat{t}\right) \hat{z}%
\right) ^{2}-\hat{P}_{z}^{2}+c\left( \hat{t}\right) \right] }\right\vert
z\right\rangle f_{2}\left( \hat{t}\right)  \notag \\
&&~=\int_{-\infty }^{+\infty }dzf_{1}\left( \hat{t}\right) \left\langle
z\left\vert e^{ia^{-1}\left( \hat{t}\right) \hat{P}_{t}\hat{P}_{z}}e^{-u%
\left[ a^{2}\left( \hat{t}\right) \hat{z}^{2}-\hat{P}_{z}^{2}+c\left( \hat{t}%
\right) \right] }e^{-ia^{-1}\left( \hat{t}\right) \hat{P}_{t}\hat{P}%
_{z}}\right\vert z\right\rangle f_{2}\left( \hat{t}\right) ,  \label{7}
\end{eqnarray}%
where the operators $\hat{P}_{z}$ and $\hat{P}_{t}$ satisfy the following
commutation relations $\left[ \hat{P}_{z},\hat{z}\right] =\left[ \hat{P}_{t},%
\hat{t}\right] =i$, and the gauge conditions%
\begin{equation*}
\left\langle z\right\vert \hat{P}_{z}\left\vert z^{\prime }\right\rangle
=-i\partial _{z}\left\langle z\right. \left\vert z^{\prime }\right\rangle
=i\partial _{z^{\prime }}\left\langle z\right. \left\vert z^{\prime
}\right\rangle ,\left\langle t\right\vert \hat{P}_{t}\left\vert t^{\prime
}\right\rangle =-i\partial _{t}\left\langle t\right. \left\vert t^{\prime
}\right\rangle =i\partial _{t^{\prime }}\left\langle t\right. \left\vert
t^{\prime }\right\rangle .
\end{equation*}%
To literally reproduce the matrix element (\ref{3}), one has to set $u=is$.
The proof of the shift theorem in \cite{CNshift06} is based on a variable
change which allows one to pass from eq. (23) to eq. (25) and then to (26).
Moreover, the transition from eq. (23) to eq. (25) is based on the
supposition that a change of the integration path over the new variable $%
z^{\prime }$ is possible. In our notation, such a variable change is $%
z=z^{\prime }+a^{-1}\left( \hat{t}\right) \hat{P}_{t}$, and it implies
introducing a new coordinate operator $\hat{z}^{\prime }=\hat{z}%
-a^{-1}\left( \hat{t}\right) \hat{P}_{t}$. However, if $a\left( t\right) $
really depends on time, then $\left[ a\left( \hat{t}\right) ,a^{-1}\left( 
\hat{t}\right) \hat{P}_{t}\right] \neq 0$ and, therefore, $\left[ a\left( 
\hat{t}\right) ,\hat{z}^{\prime }\right] \neq 0$. In addition, $z^{\prime }=$
$z-a^{-1}\left( \hat{t}\right) \hat{P}_{t}$ is not a $c$-number, since $%
\left[ a\left( \hat{t}\right) ,z^{\prime }\right] \neq 0$.

It is implicitly supposed in \cite{CNshift06} that for any $u$ (the
parameter $u$ takes all the values on the real or imaginary semiaxis) the
kernel in the integral on the left hand side (\ref{7}) has necessary
analytical properties for a change of the integration path. At the same
time, it is supposed that $\left[ a\left( \hat{t}\right) ,z^{\prime }\right]
=0$ and $z^{\prime }$ is transformed as a usual coordinate $z$ on this new
path. Rigorous justification for this is not given; only an analogy with
Gaussian integral (7) from \cite{CNshift06} is mentioned. Let us note that
in the general case, $a\left( \hat{t}\right) $ and $c\left( \hat{t}\right) $
arbitrary, the kernel on the left hand side (\ref{7}) as a function of $z$
is unknown due to the presence of the operator $\hat{P}_{z}^{2}$ in the
exponential. Thus, such an analogy does not work. We believe that such a
delusion is a result of the ambiguous notation, in which the coordinate
operator and its eigenvalue are denoted by the same letters. Therefore, the
transformation, used in \cite{CNshift06} does not exist, so that the shift
theorem does not hold. Consequently, one can believe that a number of
results obtained in \cite{CNb06,Njhep09} and based on this theorem are open
to question. In the particular case that is discussed in \cite{Nepjc09} one
has $a\left( t\right) =g\Lambda \left( t\right) $, $c\left( t\right)
=-2ig\lambda _{l}\Lambda \left( t\right) $. If the field $\Lambda \left(
t\right) $ really depends on time, which is the case, then $\left[ \Lambda
\left( t\right) ,i\Lambda ^{-1}\left( t\right) \partial _{t}\right] \neq 0$,
and the transformation from Eq.(36) to Eq.(37), given in \cite{Nepjc09},
does not exist.

\subparagraph{\protect\large Acknowledgement}

D.M.G. acknowledges the permanent support of FAPESP and CNPq.

\end{document}